\documentclass[useAMS,usenatbib]{mn2e}
\usepackage{epsfig}
\title[Polarimetric studies of comet Hale-Bopp]{Polarimetric studies of comet Hale-Bopp}
\author[H. S. Das and A. K. Sen]{H. S. Das$^{1}$\thanks{E-mail:
hs\_das@rediffmail.com (HSD); asokesen@sancharnet.in (AKS)} and A.
K. Sen$^{2}$\footnotemark[1]\\
$^{1}$Department of Physics, Kokrajhar Govt. College, Kokrajhar 783370, Assam, India\\
$^{2}$Department of Physics, Assam University, Silchar 788011,
Assam, India \\
 also Visiting Associate, IUCAA, Pune 411007,India}
\begin{document}

\date{Accepted ****. Received ****; in original form ****}

\pagerange{\pageref{firstpage}--\pageref{lastpage}} \pubyear{2007}

\maketitle

\label{firstpage}

\begin{abstract}
    In the present work, the non-spherical dust grain characteristics
    of comet Hale-Bopp
    are studied  using the T-matrix method and
    the modified power law distribution function.
    It is found that the observed data fits very well with the power
    index ($\alpha) = - 3$. The best fit values of complex refractive
    index $(n, k)$ and the aspect ratio
    (E)   at $\alpha = - 3$  are  calculated to be $(1.382, 0.035,
    0.936)$ and $(1.379, 0.041,0.936)$  at $\lambda = 0.485 \mu m$ and
    $0.684\mu m$ respectively. Kerola \& Larson (K-L) analysed the same
    comet using the T-matrix
    method and the power law distribution function ($\alpha = -3$), and
    found that the prolate grains can  explain the observed
    polarization in a more satisfactory manner as compared to the
    other shapes. But their analysis could not reproduce the negative
    polarization branch beyond scattering angle $157^0$. However, the
    results obtained from the present work successfully generate the
    expected negative polarization curve beyond $157^0$ and the
    fitting in this case is much better than K-L's work. So
    it is concluded from the present study that the use of modified power law
    distribution function (with $\alpha = - 3$) can  fit the
    observed data in a better way, as compared to the power law distribution function
    used by previous authors.
\end{abstract}


\section{Introduction}
Comets are believed to be the most primordial objects in our solar
system. They are the least processed remnants from the early solar
nebula era of dust grain formation. Thus, the study of comets
gives information about the least processed and most pristine
materials of early solar nebula, from which the present day solar
system has been evolved. The knowledge of cometary grains come
mainly from the light scattered by the grains, which are present
in the coma. The linear polarization of the scattered light by
dust particles depends upon $(i)$ wavelength of incident light
($\lambda$), $(ii)$ the shape and size  of the particle, $(iii)$
Scattering angle, $\theta$ (= $180^0 -$ Phase angle),  and $(iv)$
the composition of dust particles in terms of complex values of
refractive index, $m$ $(= n - ik)$.   For regularly shaped
spheroidal grains one can characterise their shapes by an aspect
ration E, where  grains  E $> 1$ for oblate, E $< 1$ for prolate
and E $= 1$ for spheres. However, the grains are in general not
solid but porous and irregularly shaped. In general, the study of
the polarization of continuum radiation in comets is a powerful
tool to determine the characteristics of the cometary dust grains
(Kikuchi et al. 1987; Sen et al. 1991a, 1991b; Joshi et al. 1997;
Chernova et al. 1993; Ganesh et al. 1998; Das et al. 2004 etc.).

Levasseur-Regourd et al. (1996), studied a polarimetric data base
of 22 comets and from the nature of phase angle dependence, they
concluded that there is a clear evidence for two class of comets.
These two classes of comets are distinct only for the phase angles
above 40$^0$. However, Das et al.(2004) had shown that the comets
on the basis of their dust properties need not be classified into
such discrete classes, rather their dust properties are some
smooth varying functions of  cometary  aging. Comet Hale-Bopp
C/1995 O1 is an intrinsically bright comet, with positive
polarization values much higher than those of other comets
(although the highest phase angle observed was 47$^0$). Hadamcik
\& Levasseur-Regourd (2003) compared the imaging polarimetry of
seven different comets and suggested that Hale-Bopp itself
represents a third class, marked by unusually high polarization.

It is now almost accepted  that cometary grains are not spherical
and may be \emph{fluffy aggregates} or \emph{porous}, with
irregular or spheroidal shapes (Greenberg $\&$ Hage 1990). The
measurement of circular polarization of comet Hale-Bopp (Rosenbush
et al. 1997) revealed that cometary grains are mostly
non-spherical in shape. In order to study the light scattering
properties of these  irregularly shaped  and spheroidal grains ,
Discrete Dipole Approximation (DDA) (Draine 1988)and T-matrix
theory (Waterman 1965)  are in general used respectively. Using
the DDA method, Xing $\&$ Hanner (1997) studied  the fluffy nature
of cometary grains of different shapes and sizes. Moreno et al.
(2003) studied the composite grains using the DDA method  for
modelling the comet Hale-Bopp's  grains in the mid infrared
spectrum. But, the DDA method requires considerable computer time
and memory. The T-matrix code  (Mishchenko et. al. 2002) on the
other hand runs much faster and the results obtained can be tuned
easily. Using the T-matrix code, Kerola $\&$ Larson (2001)
analysed the polarization data of comet Hale-Bopp and found the
grains to be mostly prolate in shape in that comet. Recently Das
$\&$ Sen (2006) using the T-matrix code found that, the prolate
grains can explain the observed polarization in a better way as
compared to the other shapes in comet Levy 1990XX. However, the
T-matrix code in its present form cannot be used for studying
inhomogeneous (e.g. porous, fluffy, composite) particles
(Mishchenko et al. 2002).

It is to be noted that the results obtained from the T-matrix code
could not reproduce the negative polarization branch observed for
comet Hale-Bopp, as seen in Kerola and Larson (2001), where the
analysis has been restricted to $\theta \leq 160^0$. Thus, the
main objective of the present work is to see whether a better fit
can be achieved by varying the dust size distributions and dust
shapes.

\section{Spheroidal grain model}
 As already discussed the T-matrix method provides a powerful tool to study the
 spheroidal
grains in comets. In this paper, calculation has been carried out
for randomly oriented spheroids using Mishchenko's (1998) single
scattering  T-matrix code, which is available in
\textit{http://www.giss.nasa.gov/$\sim$ crmim}.  The main feature
of the T-matrix approach is that it reduces exactly to the Mie
theory when the particle is a homogeneous or layered sphere
composed of isotropic materials.

 The \textit{in
situ} dust measurement of comet Halley gave the first direct
evidence of grain mass distribution (Mazets et al. 1986, Lamy et
al. 1987). Mukai et al. (1987) and Sen et al. (1991a) based on Mie
Theory  analysed the polarization data of comet Halley using the
power law dust distribution suggested by Mazets et al. (1986)  and
derived a set of refractive indices of cometary grains. Das et al.
(2004) also analysed the polarization data of several comets
including Halley using dust distribution function suggested by
Lamy et al. (1987). For the analysis of polarimetric data of comet
Hale-Bopp, a \textit{power law size distribution} was used by
Kerola \& Larson (K-L) (2001), where the minimum and maximum
particle radius ($r_1$ and $r_2$ respectively) are automatically
fixed for each and every run merely by specifying the particle's
effective radius ($r_{\mathrm{eff}}$) and effective variance
($v_{\mathrm{eff}}$).

This \textit{power law size distribution} (Hansen \& Travis 1974)
is given by

\begin{equation}
n(r) =
 \left\{
\begin{array}{l l}
\mathrm{constant} \times r^{-3} & , r_1 \leq r \leq r_2,\\
0 & , \mathrm{otherwise},
 \end{array}
 \right.
\end{equation}
Using equation (1) and the T-matrix code, K-L achieved reasonably
good agreement with a set of \emph{spherical volume equivalent}
values of effective radius ($r_{\mathrm{eff}}$), effective
variance ($v_{\mathrm{eff}}$) and E as 0.216 $\mu m$, 0.0105 and
0.415 respectively for prolate spheroids at $\lambda = 0.485 \mu
m$ and $0.684 \mu m$. To analyse the data, the index of refraction
for crystalline olivine (1.63, 0.00003) was taken.  However, the
analysis was restricted to $\theta \le 160^0$. The results
obtained from the above calculations could not reproduce the
negative polarization branch beyond $160^0$.

In the present study, the same comet is analysed using the
\emph{modified power law distribution function} (Mishchenko et al.
1999), which is given by

\begin{equation}
n(r) = \left\{
\begin{array}{l l}
\mathrm{constant} , & 0 \leq r \leq r_1,\\
\mathrm{constant} \times (r/r_1)^{\alpha},  & r_1 \leq r \leq r_2,\\
 0,  & r_2 < r,
 \end{array}
\right.
\end{equation}

It is  to be noted that the albedo of comet Hale-Bopp as derived
from the comparison between infrared thermal emission and visible
scattered light  seems to be higher than those for other comets
(Williams et al. 1997; Jones \& Gehrz 2000). Williams et al.
(1997) had interpreted this high albedo as an indication for the
presence of a large number of small particles in Hale-Bopp, and
Hanner et al. (1999) as an increase of the ratio between silicates
and carbonaceous compounds. Thus the higher albedo and higher
polarization could provide a clue to the presence of smaller
grains in comet Hale-Bopp (Hadamcik \& Levasseur-Regourd 2003).

 In the present work, $r_1$ and $r_2$ are taken to be $0.01 \mu m$ and
$2 \mu m$ respectively and a few early iterations using
contrasting values of $\alpha$ are tried. Initially, an index of
refraction for crystalline olivine (1.63, 0.00003) is taken here
as fixed for the analysis of data. Thus the shape parameter ($E$)
was left as the only free parameter to vary.   Taking $\alpha =
-3$, the shape parameter ($E$) is varied to fit the observed data
at $\lambda = 0.485 \mu m$ by $\chi ^2$ minimisation technique.
But no good fit was observed at $\alpha = -3$. E is varied for
other values of $\alpha$ (say, -1.1, -1.2, -1.3,..., -2.9, -3.0,
-3.1 etc.), but none of them could match the observed data well.

So a different approach is proposed here. The refractive index
parameter $(n, k)$ is now taken as another free parameter. Taking
a particular value of $\alpha$, the best fit values of $(n, k)$
and $E$ are determined at which the the sum of squares of
differences between the calculated and observed values of
polarization ($\chi ^2$ - value) becomes minimum. This calculation
is repeated for several other values of $\alpha$. The results
obtained from the present work are reproduced in \textbf{Table 1}.
It can be seen from \textbf{Table 1} that a value of $\alpha = -3$
fits the data well both at $0.485 \mu m$ and $0.684\mu m$ and the
best fit values of  $(n, k)$ and E are estimated to be $(1.382,
0.035, 0.936)$ and $(1.379, 0.041,0.936)$ at $\lambda = 0.485 \mu
m$ and $0.684\mu m$ respectively. The $\chi ^2_{\mathrm{min}}$ -
values for spherical grains ($E=1$) are also shown in
\textbf{Table 1}. Thus one can see that the prolate grains
($E=0.936$) represent a better fit to the observed data with lower
$\chi^2_{\mathrm{min}}$ value, as compared to spherical grains.
Moreover, the uniqueness of the estimated $E$ and $\alpha$ values
at two different wavelengths, further strengthens our claim for a
more suitable and realistic fit to the observed data.

Using Mie scattering theory and grain model of Mazets et al.
(1986), Mukai et al. (1987) analysed comet Halley and found a set
of three complex refractive indices $(n, k)$ at three IHW filters
which best matched their observations. Sen et al. (1991a) combined
their polarimetric observations with those of other investigators
and estimated $(n, k)$ values which are slightly different from
those  of Mukai et al. (1987). Based on the dust size distribution
function derived by Das et al. (2004) for comet Halley (on the
basis of the work reported by Lamy et al. (1987)) and Mie theory,
Das et al. (2004) also analysed polarization  data and found a set
of refractive indices $(n, k)$ for comet Halley.  The best fit
values of $(n, k)$ derived by them for comet Halley are reproduced
in \textbf{Table 2}.  Lamy et al. (1987) denoted these
hypothetical refractive indices $(n, k)$ emerging out from the Mie
code as `Silicate B'. The present analysis also suggests the
`Silicate B' nature of  comet Hale-Bopp's dust grains.

The $\chi ^2_{\mathrm{min}}$ - values obtained from present work
are 20.8 and 39.1 at $\lambda = 0.485 \mu m$ and $0.684\mu m$
respectively, whereas the values obtained from K-L's work are
212.9 and 174.4 respectively.  Thus the present analysis is
clearly giving better fit to the observed polarization data of
comet Hale-Bopp, as compared to K-L.  In \textbf{Fig 1} and
\textbf{Fig 2}, the best fitted polarization curves obtained from
the T-matrix code are drawn on the observed polarization data
(Ganesh et al. 1998, Manset \& Bastien 2000) at $\lambda = 0.485
\mu m$ and $0.684\mu m$ respectively. It is interesting to note
that the present analysis can reproduce the negative polarization
curves beyond $157^0$  at $\lambda = 0.485 \mu m$ and $0.684\mu m$
respectively, which were not possible in K-L's work. The simulated
polarization curves from K-L's work are also shown in \textbf{Fig
1} and \textbf{Fig 2}.

\begin{table*}
%\begin{center}
 \caption{The best fit values of $(n, k)$  and E obtained in the present
work for comet Hale-Bopp at $\alpha = -2.9, -3.0$ and $-3.1$.}
\begin{tabular}{|c|c|c|c|c|c|c|c|}
\hline
  $\lambda$   & Scattering angle& No. of data & $\alpha$&$n$&$k$ & E& $\chi
  ^2_{\mathrm{min}}$  \\
(in $\mu m$) &range (in degrees)  &points &&&&&\\
  \hline
 & & & -2.9 &1.381&0.039 &0.936 &22.13\\
\cline{7-8}
 & & & & & &1.000&73.21\\
\cline{4-8}
0.485&133 - 163&29& -3.0 &1.382&0.035&0.936&20.84\\
  \cline{7-8}
 & & & & & &1.000&68.36\\
\cline{4-8}
 & & & -3.1 &1.383&0.031 &0.936 &22.16\\
\cline{7-8}
 & & & & & &1.000&52.65\\
  \cline{1-8}
  & & & -2.9 &1.379 &0.046 &0.936 &40.18\\
\cline{7-8}
 & & & & & &1.000&89.44\\
\cline{4-8}
  0.684&133 - 177&57&-3.0 &1.379&0.041&0.936&39.08\\
  \cline{7-8}
 & & & & & &1.000&89.82\\
\cline{4-8}
 & & & -3.1 &1.379 &0.036 &0.936 &41.10\\
\cline{7-8}
 & & & & & &1.000&87.05\\
  \hline
\end{tabular}
%\end{center}
\end{table*}
%%%%%%%%%%%%%%%%%%%%%%%%%%%%%%%%%%%%%%%%%%%%%%%%%%%%%%%%%%%%%%%%%%%%
\begin{table*}
\begin{center}
\caption{The $(n, k)$ values as obtained by different authors  for
comet Halley using the Mie code at  different wavelengths.}
\begin{tabular}{|c|c|c|c|}
\hline

\bf $\lambda$ & $n$ & $k$ & Authors\\

\hline

  & 1.392 & 0.024 & Mukai et al. (1987)  \\
\cline{2-4}
0.365 $\mu m$& 1.387& 0.032& Sen et al. (1991a)\\
\cline{2-4}
 & 1.403& 0.024 & Das et al. (2004)\\
\hline
  &1.387 &0.031 & Mukai et al. (1987)\\
\cline{2-4}
0.485 $\mu m$& 1.375& 0.040& Sen et al. (1991a)\\
\cline{2-4}
  &1.390 &0.026 & Das et al. (2004)\\
\hline
0.620 $\mu m$ & 1.385 &0.035 & Mukai et al. (1987)\\
\hline
0.684 $\mu m$ &1.374 &0.052 & Sen et al. (1991a)\\
\cline{2-4}
 & 1.386 &0.038 & Das et al. (2004)\\
 \hline

\end{tabular}
\end{center}
\end{table*}

    \begin{figure*}
   \centering
   \includegraphics{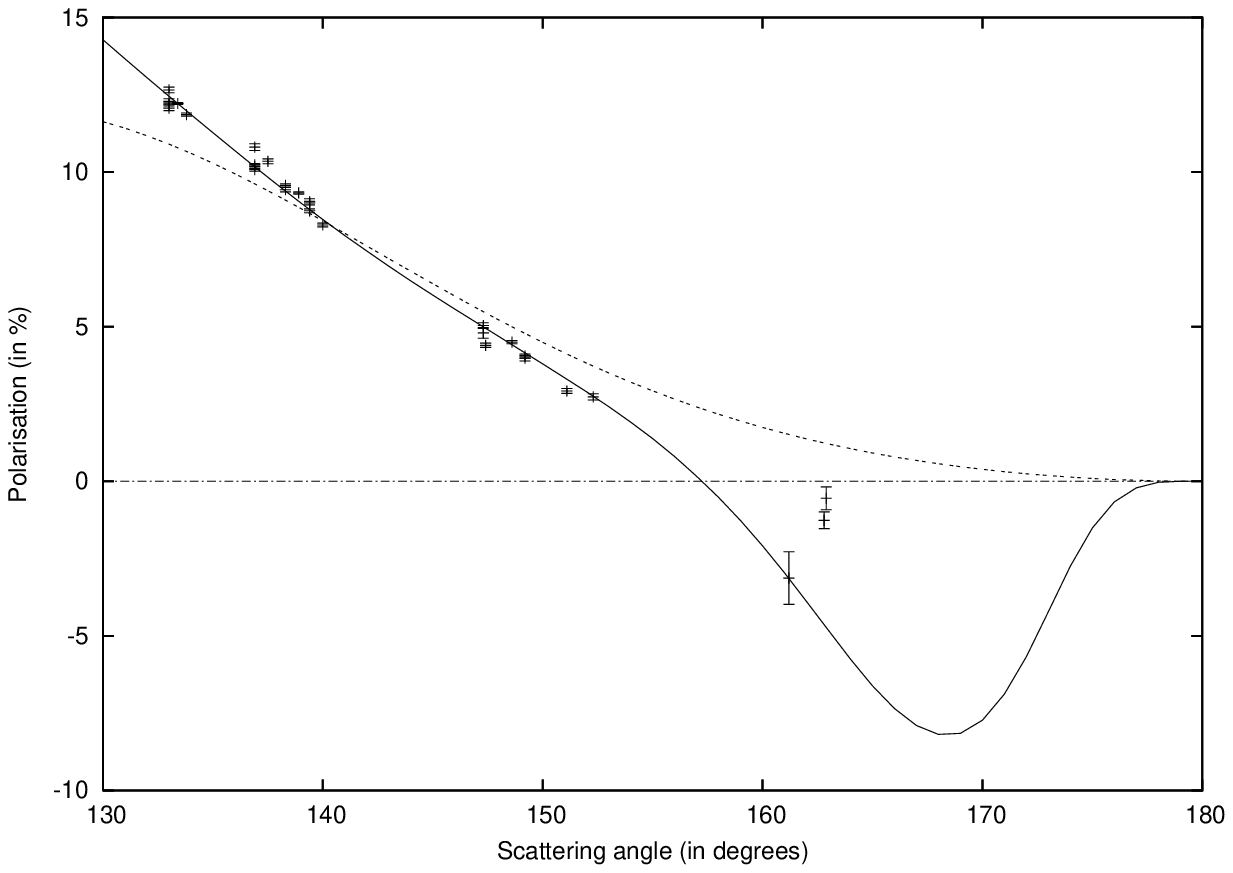}
   \caption{ The observed polariation values of comet Hale-Bopp at
   $\lambda =0.485 \mu m$. The solid line represents the best fitted polarization
   curve obtained from the present work. The simulated polarization curve
   from  Kerola \& Larson (2001) is  also shown by dotted line. }
    \end{figure*}

 \begin{figure*}
   \centering
   \includegraphics{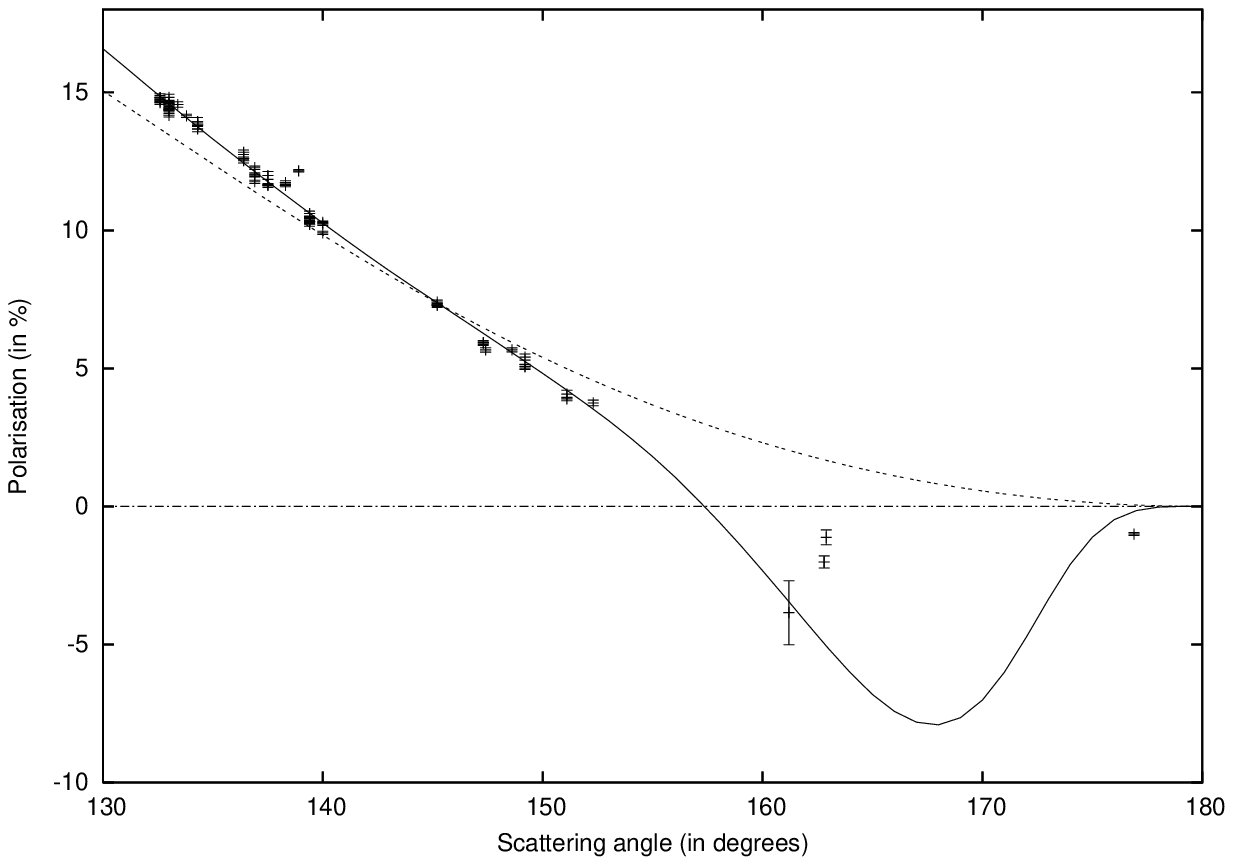}
   \caption{The observed polariation values of comet Hale-Bopp at
   $\lambda =0.684 \mu m$. The solid line represents the best fitted polarization
   curve obtained from the present work. The simulated polarization curve
   from Kerola \& Larson (2001) is  also shown by dotted line.  }
    \end{figure*}

\section{Discussions}
It can be seen from the above analysis that the simulated
polarization values in the present case fit much better to the
observed data as compared to K-L's work. The main reasons for
getting better fit are : $(i)$ use of modified power law
distribution (though the power index is same in both the
distributions), $(ii)$ the size range of the grains, which is
slightly broader in the present work as compared to K-L's work,
$(iii)$ the refractive index  $(n, k)$ in the present analysis,
which is also different from K-L's work where the index of
refraction for crystalline olivine was considered and finally
$(iv)$ the shape parameter which is $E=0.936$ in the present work,
as against the value $E=0.415$ used by K-L.

In the present study, one can generate the expected polarization
curves beyond $157^0$ at $\lambda = 0.485 \mu m$ and $0.684\mu m$.
But, the results obtained by K-L  could not reproduce the negative
polarization branch for comet Hale-Bopp beyond $157^0$. Their
analysis was restricted to $\theta \leq 160 ^0$. K-L concluded
that combination of viewing geometry effects and enhanced multiple
scattering might provide a quantitative explanation of the
negative polarization for scattering angle beyond $160^0$.

Greenberg \& Hage (1990) originally proposed the presence of large
numbers of \emph{porous} grains in the coma of comets to explain
the spectral emission at 3.4 $\mu m$ and 9.7 $\mu m$. Dollfus
(1989) discussed the results of laboratory experiments by
microwave simulation and laser scattering on various complex
shapes with different porosities. It has been observed that the
measurement of 10 $\mu m$ flux in comet Hale-Bopp is dominated by
small,  porous, amorphous silicate grains (Harker et al. 2002).
The higher polarization observed for the dust coma of comet
Hale-Bopp also suggests the existence of smaller grains included
in highly porous large aggregates, and possibly an increase in the
number of high albedo grains in these aggregates (Hadamcik \&
Levasseur-Regourd 2003). The \textit{fluffy aggregate model}
originally proposed by Greenberg and Hage (1990) and later adopted
by Xing and Hanner (1997) is also used for the study of negative
polarization in comets. Tanga et al. (1997) and Levasseur-Regourd
et al. (1998) suggested that \emph{multiple scattering} may well
explain the negative polarization, because lower polarization is
found in the near-nucleus region of comets where dusty jets are
most pronounced. Xing \& Hanner (1997) have done calculations with
porous grains of various shapes and sizes using the DDA method.
Petrova et al. (2000) have shown that aggregates composed of
touching spheres ($< 50 \mu m$) with size parameters ranging from
1.3 to 1.65, display properties typical of cometary dust
particles, namely, a weak increase of the backscattering
intensity, negative linear polarization at small phase angles
($\le 20^0$), and a positive wavelength gradient of polarization.
Their results on the aggregates indicate that more compact
particles have a more pronounced negative branch of polarization.
They also commented that the increase of polarization with
wavelength is reduced, if the imaginary part of the refractive
index decreases with wavelength.  Moreno et al. (2003) also
studied the irregular and composite grain characteristics of comet
Hale-Bopp in the mid infrared spectrum using the DDA method.  The
favoured grain model is now that of porous \emph{fluffy} grains
with irregular shapes. So, it is important to study the
\emph{fluffy} grains with irregular shapes and enhanced multiple
scattering, which may better explain the observed polarization
data in comets. However, a systematic approach in this direction
is beyond the scope of the present work.

\section{Conclusions}
Based on the  T-matrix method and the modified power law
distribution function, the following conclusions can be drawn from
the present work:
   \begin{enumerate}
\item The complex refractive indices and shape parameters of
Hale-Bopp's grains as derived from present work are: (1.382,
0.035, 0.936), (1.379, 0.041, 0.936) at $\lambda =  0.485 \mu m$
and $0.684\mu m$ respectively.

\item It has been found from the present analysis that
the power  index  ($\alpha$) in the modified power law
distribution is  same as that of the power law distribution
function used by K-L, which is -3.

\item The $\chi ^2_{\mathrm{min}}$ - values obtained from present work
are 20.84 and 39.08 at $\lambda = 0.485 \mu m$ and $0.684\mu m$
respectively, whereas the corresponding values from K-L's work are
are 212.9 and 174.4 respectively.  Thus the present analysis is
giving better fit to the observed polarization data of comet
Hale-Bopp as compared to K-L.

\item The expected negative polarization values have been
successfully generated for comet Hale-Bopp using the T-matrix
method, which was not possible earlier by K-L.

\end{enumerate}

 \section{Acknowledgements}
 The authors sincerely acknowledge IUCAA, Pune, where some part of
 these calculations were done. The authors are  also thankful to M.
 Mishchenko for the T-matrix code. Finally the author AKS is thankful to
 Indian Space Research Organisation (ISRO) for funding the RESPOND project
 (sanction no. 9/2/92/2004-II dtd. Dec 21, 2004) under which the work was done.

\end{document}